\documentclass[runningheads]{llncs}

\usepackage[T1]{fontenc}
\usepackage{amsmath}
\usepackage{amssymb}
\usepackage{graphicx}
\usepackage{multirow}
\usepackage{booktabs}
\usepackage{array}
\usepackage{url}
\usepackage{tikz}
\usetikzlibrary{arrows.meta,positioning}
\usepackage{pgfplots}
\pgfplotsset{compat=1.17}
\usepackage{float}
\usepackage[linesnumbered,ruled,vlined]{algorithm2e}
\SetAlgoNlRelativeSize{0}
\definecolor{trainone}{HTML}{1A5276}   
\definecolor{traintwo}{HTML}{A93226}   
\definecolor{conflict}{HTML}{E67E22}   
\tikzset{
  t1/.style={draw=trainone!85!black, fill=trainone!10!white, rounded corners=2pt, thick, align=center},
  t2/.style={draw=traintwo!85!black, fill=traintwo!10!white, rounded corners=2pt, thick, align=center},
  t2ghost/.style={draw=traintwo!50, fill=traintwo!2!white, dotted, thick, rounded corners=2pt},
  ovl/.style={fill=conflict, fill opacity=0.25, draw=conflict!85!black, dashed, thick, rounded corners=2pt},
  seg/.style={anchor=east, font=\bfseries\sffamily\large, text=gray!80},
  tlabel/.style={font=\small\bfseries\sffamily},
}

\let\oldbibliography\bibliography
\renewcommand{\bibliography}[1]{%
  \let\oldthebibliography\thebibliography
  \renewcommand{\thebibliography}[1]{%
    \oldthebibliography{##1}%
    \setlength{\itemsep}{0pt plus 0.2pt}%
    \setlength{\parsep}{0pt}%
  }%
  \oldbibliography{#1}%
}

\begin{document}
\raggedbottom

\title{An Efficient MaxSAT-DDD Approach for Train Rescheduling via Precedence Propagation and Hybrid AMO Encodings}
\titlerunning{An Efficient MaxSAT-DDD Approach for Train Rescheduling}

\author{Tuyen Van Kieu\inst{1} \and
Tan Huu Nguyen\inst{1} \and
Khanh Van To\inst{1}\thanks{Corresponding author.}}
\authorrunning{T. V. Kieu et al.}
\institute{Faculty of Information Technology, VNU University of Engineering and Technology, Hanoi, Vietnam\\
\email{tuyenkv@vnu.edu.vn} \quad
\email{22028102@vnu.edu.vn} \quad
\email{khanhtv@vnu.edu.vn}}

\maketitle

\begin{abstract}
Train rescheduling repairs disturbed timetables while enforcing train-path
precedence, resource capacity, and delay objectives. Dynamic Discretization
Discovery (DDD) avoids full time discretization by refining only time points
needed to certify feasibility and optimality. We strengthen a recent
MaxSAT-DDD model through two encoding changes. First, resource conflicts are
encoded as time-dependent at-most-one cliques, using pairwise clauses for small
cliques and a sequential counter for large cliques. Second, earliest feasible
times are propagated along train paths before the first DDD iteration. We
evaluate four MaxSAT variants, two SAT optimization backends, Gurobi/CPLEX
MILP models, and CPLEX CP on 72 instances and three delay objectives.
MaxSAT-DDD solves all stepwise instances in about 23 ms on average.
MaxSAT-Default reduces rounded-cost runtime from 794 to 479 ms, and the
ablation study reports up to 79.6\% runtime reduction on the common-solved subset of hard continuous track
instances.

\keywords{Train rescheduling \and Dynamic Discretization Discovery \and MaxSAT \and SAT encoding \and Sequential counter \and Precedence propagation}
\end{abstract}

\section{Introduction and Related Work}

Railway traffic management must react to delays, disrupted rolling stock, or
temporary capacity reductions while keeping the resulting plan safe and close to
the published timetable. In fixed-route train rescheduling, the dispatcher
chooses entry times for resources such as block sections, tracks, and stations.
This fixed-route assumption is a modeling choice that simplifies the scheduling search space; however, it limits applicability to minor delay recovery scenarios rather than large-scale disruptions where train rerouting, cancellations, or short-turnings are operationally required.
The model contains train-path precedence and disjunctive safety constraints,
which makes exact optimization difficult under real-time limits
\cite{FangYao2015,Cacchiani2014,Lamorgese2018}.

MILP formulations are the most common exact modelling tool for train
rescheduling. Big-$M$ formulations keep continuous time variables, but their
linear relaxations are often weak. Time-indexed formulations replace temporal
disjunctions by binary choices over a discretized horizon and usually give
stronger combinatorial structure, but a full discretization can be too large and
a coarse discretization can produce modelling errors. Dynamic Discretization
Discovery (DDD) addresses this tension by solving restricted time-indexed
models and refining only time points that violate the continuous-time
constraints \cite{Boland2017}.

Croella et al.~\cite{Croella2024} encoded restricted DDD subproblems as
weighted partial MaxSAT instances, with hard clauses for feasibility and soft
clauses for delay cost. We keep that DDD proof structure and target two CNF
bottlenecks that become visible under track and station aggregation. Aggregated
resources create large conflict cliques, and pairwise exclusions then create
$\Theta(n^2)$ clauses. Initial lower-bound ladders also ignore path precedence,
so early DDD iterations rediscover constraints already implied by each train
route.

This paper develops a more structured MaxSAT-DDD encoding. The novelty is not a
new DDD algorithm, but the integration of clique-based cardinality encoding and
train-path bound propagation into the MaxSAT-DDD refinement loop, together with
an empirical study of their behaviour under resource aggregation. The contributions
are as follows.

\begin{itemize}
\item We encode resource conflicts as time-dependent occupation cliques and use
a hybrid AMO encoding. Large cliques use the sequential-counter encoding of
Sinz~\cite{Sinz2005}.
\item We propagate effective earliest times along each train path before the
first DDD iteration. This tightening is linear in the total route length and
preserves all feasible schedules.
\item We compare four MaxSAT variants, two SAT optimization backends, Gurobi
and CPLEX MILP models, and a CPLEX CP model. We report feasible incumbents and
proven optima separately.
\end{itemize}

The results show that MaxSAT-DDD is most effective for discrete delay
objectives. Commercial Big-$M$ MILP remains strongest on continuous delay, but
MaxSAT-Default improves over the baseline MaxSAT encoding on rounded costs and
on hard continuous subsets.

Table~\ref{tab:related-work} classifies the main lines of railway
rescheduling research. The surveys in
\cite{FangYao2015,Cacchiani2014,Zhan2024Handling} show that modelling choices
depend on the operational scope, from microscopic safety constraints to
large-network disruption management.

\begin{table}
\centering
\caption{Classification of representative studies related to railway rescheduling.}
\label{tab:related-work}
\scriptsize
\setlength{\tabcolsep}{2.7pt}
\renewcommand{\arraystretch}{1.0}
\begin{tabular}{>{\raggedright\arraybackslash}p{24mm}>{\raggedright\arraybackslash}p{11mm}>{\raggedright\arraybackslash}p{25mm}>{\raggedright\arraybackslash}p{17mm}>{\raggedright\arraybackslash}p{35mm}}
\toprule
Research stream & Key studies & Main technique & Optimality role & Relation to this paper \\
\midrule
Surveys and recovery frameworks &
\cite{FangYao2015,Cacchiani2014,Lamorgese2018,Zhan2024Handling} &
Taxonomies of rescheduling models, disruption types, and recovery objectives &
Survey &
Motivates fixed-route real-time TRP and the need to report feasibility and proof status separately. \\
Continuous-time MILP &
\cite{LamorgeseMannino2015,ManninoNakkerud2023,Zhang2023Train} &
Big-$M$ disjunctions, decomposition, branch-and-cut &
Exact when proved &
Provides the strongest baseline for continuous delay objectives in our experiments. \\
Time-indexed and dynamic discretization &
\cite{Caimi2012,Boland2017,Croella2024} &
Discrete time variables, restricted time grids, iterative refinement &
Exact under DDD termination &
Forms the modelling foundation; we improve the Boolean encoding of the restricted DDD model. \\
Heuristics and simulation-optimization &
\cite{Sama2017,Bettinelli2017,Hassannayebi2020Simulation-optimization,Su2023Integrated} &
VNS, real-time conflict resolution, simulation, ALNS hybrids &
Incumbent-focused &
Targets fast operational decisions, but does not provide the optimality certificate required here. \\
Logic-based optimization &
\cite{Matos2021,Abels2021,LeutwilerCorman2022,Croella2024} &
SAT, ASP, logic-based Benders, weighted partial MaxSAT &
Exact or proof-oriented &
Closest line of work; our contribution is CNF-level engineering for MaxSAT-DDD. \\
Learning-based rescheduling &
\cite{Cao2022Train,Dong2024Deep,Liao2021A,Kovari2023Multi-Agent,Huang2025,Sehmisch2025A} &
DRL/MARL, graph neural networks, data-driven problem reduction &
Policy or warm-start &
Complementary to exact DDD; learned decisions may later guide refinement or warm starts. \\
\bottomrule
\end{tabular}
\end{table}

The closest line is logic-based exact optimization. SAT, ASP, and MaxSAT
methods exploit the Boolean structure of restricted time-indexed models
\cite{Matos2021,Abels2021,LeutwilerCorman2022}. Our work keeps the DDD
optimality logic of \cite{Croella2024} and improves its generated CNF through
sequential-counter resource cliques and precedence propagation.

\section{Train Rescheduling Problem Formulation}

Let $I$ be the set of trains. Each train $i\in I$ follows a fixed ordered route
$R_i=\langle r^i_1,\ldots,r^i_{|R_i|}\rangle$ through railway resources. The
decision variable $t^{ir}$ is the time at which train $i$ enters resource
$r\in R_i$. The input provides an earliest local time $\underline t^{ir}$, a
minimum traversal or dwell time $l_i^r$, and safety separations
$l_{ij}^{rq}$ for pairs of operations that cannot overlap. The planning horizon
$H$ bounds all entry times. Let $a_i$ be the scheduled arrival time of train $i$
at its destination $r_i^{\mathrm{dest}}$, and define
$\delta_i=\max\{0,t^{i r_i^{\mathrm{dest}}}-a_i\}$.

A schedule is feasible when it satisfies the following constraints:
\begin{align}
t^{ir} &\ge \underline t^{ir},
&& i\in I,\ r\in R_i, \label{eq:lb}\\
t^{iq} &\ge t^{ir}+l_i^r,
&& r \prec_i q, \label{eq:path}\\
t^{jq} &\ge t^{ir}+l_{ij}^{rq}
\quad \text{or}\quad
t^{ir}\ge t^{jq}+l_{ji}^{qr},
&& (r,q)\in\mathcal D_{ij}. \label{eq:conflict}
\end{align}
Constraint \eqref{eq:lb} enforces release times, \eqref{eq:path} enforces the
order of operations on a train path, and \eqref{eq:conflict} enforces resource
safety between trains. Separations $l_{ij}^{rq}$ are computed from signaling rules: under fixed-block signaling, headway limits ensure train $j$ cannot enter block $q$ until train $i$ clears $r$ plus a safety margin; at stations, they represent route locking/releasing times. Traversal times $l_i^r$ depend on speed limits and train dynamics. The objective is $\min \sum_i c_i(\delta_i)$, with three delay-cost functions:
\begin{itemize}
\item \emph{Step}: $c_i(\delta)=\sum_k w_{ik}\mathbf 1[\delta>\eta_{ik}]$.
\item \emph{Round}: $c_i(\delta)=w_i\lceil\delta/180\rceil$, where the 180-minute granularity is inherited from the benchmarks of \cite{Croella2024} (representing coarse step-like penalties).
\item \emph{Cont}: $c_i(\delta)=w_i\delta$.
\end{itemize}

\subsection{Illustrative Conflict-Resolution Example}
\begin{figure}[!htbp]
\centering
\begin{tikzpicture}[xscale=0.7, yscale=0.55]
  \def\h{0.7}
  \foreach \y/\name in {3.6/$a$, 2.4/$b$, 1.2/$c$, 0/$d$} {
    \draw[gray!25, thin] (0,\y) -- (14.5,\y);
    \node[seg] at (-0.3, {\y+0.35}) {\name};
  }
  \draw[->, thick] (0,-0.7) -- (15,-0.7) node[right, font=\small\sffamily] {Time};
  \foreach \t in {0,2,4,6,8,10,12,14} {
    \draw[thin] (\t,-0.6) -- (\t,-0.8) node[below, font=\small\sffamily] {\t};
  }
  
  \draw[t1] (0,3.6) rectangle (3,{3.6+\h});
  \node[trainone, tlabel] at (1.5, {3.6+0.35}) {T1};
  
  \draw[t2] (0,2.4) rectangle (5,{2.4+\h});
  \node[traintwo, tlabel] at (2.5, {2.4+0.35}) {T2};
  
  \draw[t1] (3,1.2) rectangle (7,{1.2+\h});
  \draw[t2] (5,1.2) rectangle (9,{1.2+\h});
  \filldraw[ovl] (5,1.2) rectangle (7,{1.2+\h});
  \node[trainone, tlabel] at (4, {1.2+0.35}) {T1};
  \node[traintwo, tlabel] at (8, {1.2+0.35}) {T2};
  
  \draw[t1] (7,0) rectangle (10,\h);
  \draw[t2] (9,0) rectangle (12,\h);
  \filldraw[ovl] (9,0) rectangle (10,\h);
  \node[trainone, tlabel] at (8, 0.35) {T1};
  \node[traintwo, tlabel] at (11, 0.35) {T2};
  
  \node[conflict, font=\footnotesize\bfseries\sffamily] at (6, {1.2+\h+0.28}) {conflict};
  \node[conflict, font=\footnotesize\bfseries\sffamily] at (9.5, {\h+0.28}) {conflict};
  
  \draw[t1] (1,-2.1) rectangle (1.8,-1.7);
  \node[anchor=west, font=\footnotesize\sffamily] at (1.9,-2.0) {Train 1};
  \draw[t2] (4,-2.1) rectangle (4.8,-1.7);
  \node[anchor=west, font=\footnotesize\sffamily] at (4.9,-2.0) {Train 2};
  \filldraw[ovl] (7,-2.1) rectangle (7.8,-1.7);
  \node[anchor=west, font=\footnotesize\sffamily] at (7.9,-2.0) {Conflict region};
  
  \node[anchor=north, font=\small\bfseries\sffamily] at (6.5, -2.4)
  {(a) Schedule with initial conflicts};
\end{tikzpicture}

\vspace{0.1cm}

\begin{tikzpicture}[xscale=0.7, yscale=0.55]
  \def\h{0.7}
  \foreach \y/\name in {3.6/$a$, 2.4/$b$, 1.2/$c$, 0/$d$} {
    \draw[gray!25, thin] (0,\y) -- (14.5,\y);
    \node[seg] at (-0.3, {\y+0.35}) {\name};
  }
  \draw[->, thick] (0,-0.7) -- (15,-0.7) node[right, font=\small\sffamily] {Time};
  \foreach \t in {0,2,4,6,8,10,12,14} {
    \draw[thin] (\t,-0.6) -- (\t,-0.8) node[below, font=\small\sffamily] {\t};
  }
  
  \draw[t1] (0,3.6) rectangle (3,{3.6+\h});
  \node[trainone, tlabel] at (1.5, {3.6+0.35}) {T1};
  
  \draw[t2] (0,2.4) rectangle (5,{2.4+\h});
  \node[traintwo, tlabel] at (2.5, {2.4+0.35}) {T2};
  
  \draw[t2ghost] (5,1.2) rectangle (9,{1.2+\h});
  \draw[t1] (3,1.2) rectangle (7,{1.2+\h});
  \draw[t2] (7,1.2) rectangle (11,{1.2+\h});
  \node[trainone, tlabel] at (4, {1.2+0.35}) {T1};
  \node[traintwo, tlabel] at (10, {1.2+0.35}) {T2};
  \draw[<->, traintwo!90!black, thick] (5,{1.2-0.45}) -- (7,{1.2-0.45})
       node[midway, below, font=\footnotesize\bfseries\sffamily] {delay +2};
       
  \draw[t1] (7,0) rectangle (10,\h);
  \draw[t2] (11,0) rectangle (14,\h);
  \node[trainone, tlabel] at (8.5, 0.35) {T1};
  \node[traintwo, tlabel] at (12.5, 0.35) {T2};
  
  \draw[t1] (1,-2.1) rectangle (1.8,-1.7);
  \node[anchor=west, font=\footnotesize\sffamily] at (1.9,-2.0) {Train 1};
  \draw[t2] (4,-2.1) rectangle (4.8,-1.7);
  \node[anchor=west, font=\footnotesize\sffamily] at (4.9,-2.0) {Train 2};
  \draw[t2ghost] (7.3,-2.1) rectangle (8.1,-1.7);
  \node[anchor=west, font=\footnotesize\sffamily] at (8.2,-2.0) {Initial planned position};
  
  \node[anchor=north, font=\small\bfseries\sffamily] at (6.5, -2.4)
  {(b) Rescheduled conflict-free timetable};
\end{tikzpicture}
\caption{An illustrative example of the train rescheduling problem (TRP), with time in minutes.}
\label{fig:trp-example}
\end{figure}
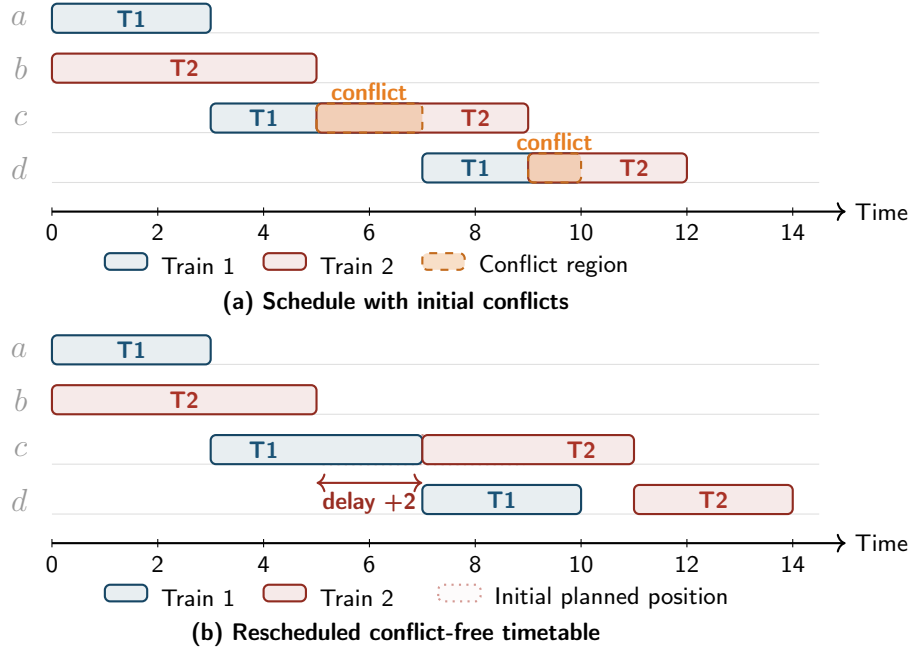

Figure~\ref{fig:trp-example} shows two fixed-route trains with shared segments
$c$ and $d$. Train 1 follows $\langle a,c,d\rangle$ and Train 2 follows
$\langle b,c,d\rangle$, with traversal times of $3, 5, 4, 3$ minutes on segments $a,b,c,d$, respectively.

In Fig.~\ref{fig:trp-example}(a), the trains overlap on $c$ and $d$, violating
\eqref{eq:conflict}. Delaying Train 2 by two time units removes both conflicts
and creates the rescheduled timetable in Fig.~\ref{fig:trp-example}(b).

\section{Baseline MaxSAT-DDD Formulation}

DDD stores a time partition $\Lambda^{ir}$ for each operation $(i,r)$ and
solves a restricted interval assignment problem. The selected interval
left-endpoints define a lower-bound schedule. Violated path or resource
constraints trigger interval refinement. If the decoded schedule is feasible
and has the same cost as the restricted model, optimality is certified.

MaxSAT-DDD encodes each restricted model as weighted partial MaxSAT. Hard
clauses enforce interval assignment, monotonicity, and discovered
incompatibilities. Soft clauses encode delay costs. Ladder literals
$d^{ir}(\tau)$ represent $t^{ir}\ge\tau$, and baseline resource conflicts are
mostly pairwise exclusions over occupation events.
For each destination-delay breakpoint $\tau$, the soft literal
$\neg d^{i r_i^{\mathrm{dest}}}(a_i+\tau)$ carries the marginal cost increase
$c_i(\tau)-c_i(\tau^-)$. Step and rounded objectives have few breakpoints,
whereas the continuous objective can create many unit-weight levels.

\section{Proposed MaxSAT-DDD Encoding}

\subsection{Solver Architecture and Refinement Workflow}

The solver first applies precedence propagation to tighten the initial ladders.
The DDD loop then alternates between optimizing the restricted MaxSAT model and
adding path, resource, or cost-refinement clauses until feasibility and objective
consistency certify optimality.

\begin{figure}[!htbp]
\centering
\begin{tikzpicture}[
  node distance=2.6mm,
  box/.style={draw=black, thick, rounded corners=1.5pt, fill=white,
    align=center, text width=18mm, minimum height=5.6mm, font=\tiny\sffamily},
  arr/.style={-{Stealth[length=1.4mm]}, draw=black, semithick},
  looparr/.style={-{Stealth[length=1.4mm]}, draw=black, dashed, semithick},
  lab/.style={font=\tiny\sffamily, fill=white, inner sep=0.8pt}
]
\node[box] (input) {TRP\\instance};
\node[box, right=of input] (prec) {Precedence\\bounds};
\node[box, right=of prec] (loop) {Restricted\\MaxSAT-DDD};
\node[box, right=of loop] (out) {Schedule\\or bounds};
\draw[arr] (input) -- (prec);
\draw[arr] (prec) -- (loop);
\draw[arr] (loop) -- (out);
\draw[looparr] (loop.south east) to[out=320,in=220,looseness=1.15]
  node[below, lab] {path/resource/cost refinement} (loop.south west);
\end{tikzpicture}
\caption{Architecture of the strengthened MaxSAT-DDD solver.}
\label{fig:architecture}
\end{figure}
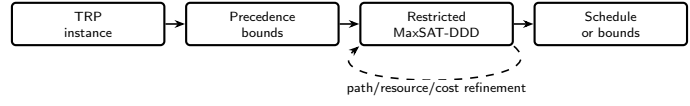

\subsection{Train-Path Precedence Propagation}

The baseline initializes each operation from its local lower bound
$\underline t^{ir}$. For route
$R_i=\langle r^i_1,\ldots,r^i_{|R_i|}\rangle$, we instead compute effective
earliest times by
\begin{align}
\underline t^{i r^i_1}_{\mathrm{eff}} &=
\underline t^{i r^i_1}, \label{eq:effbase}\\
\underline t^{i r^i_{k+1}}_{\mathrm{eff}} &=
\max\left\{\underline t^{i r^i_{k+1}},
\underline t^{i r^i_k}_{\mathrm{eff}}+l_i^{r^i_k}\right\}.
\label{eq:effrec}
\end{align}
This forward pass costs $O(\sum_i |R_i|)$ and removes no feasible schedule,
because every solution satisfying \eqref{eq:lb} and \eqref{eq:path} also
satisfies $t^{ir}\ge\underline t^{ir}_{\mathrm{eff}}$.

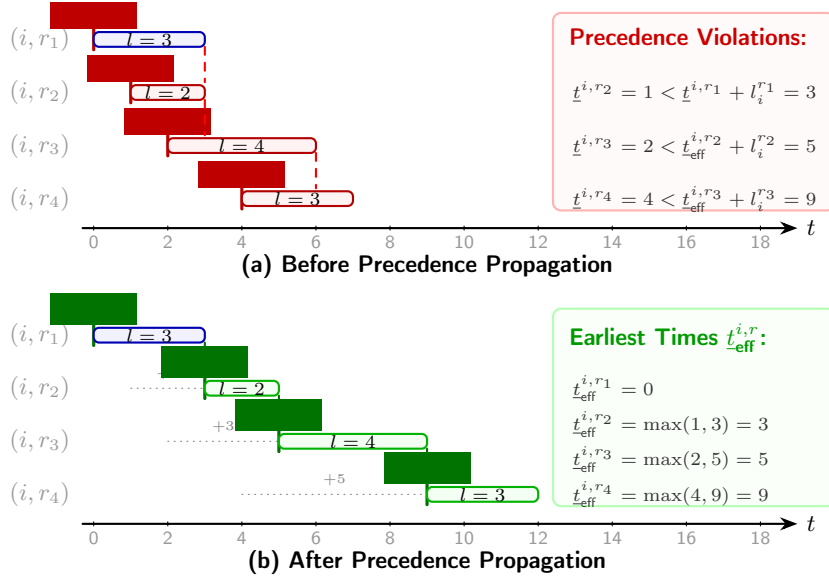
\begin{figure}[!htbp]
\centering

\begin{tikzpicture}[
    x=4.9mm, y=3.9mm, >=Stealth, line cap=round,
    chang/.style    ={draw=blue!70!black, fill=blue!5, rounded corners=2pt, thick, minimum height=4mm, inner sep=1pt},
    changwarn/.style={draw=red!70!black, fill=red!5, rounded corners=2pt, thick, minimum height=4mm, inner sep=1pt},
    lbmark/.style   ={red!75!black, line width=1.2pt},
    tlab/.style     ={font=\scriptsize\sffamily, fill=white, inner sep=1pt},
    vname/.style    ={font=\small\bfseries\sffamily, anchor=east, text=gray!80},
    ax/.style       ={font=\scriptsize\sffamily, text=gray!80}
]
\draw[->, thick] (-0.3, 0) -- (19.0, 0) node[right, font=\small\sffamily] {$t$};
\foreach \t in {0,2,4,6,8,10,12,14,16,18}
  \draw (\t, 0.07) -- (\t, -0.07) node[below=-1pt, ax] {\t};

\node[vname] at (-0.4, 6.4) {$(i, r_1)$};
\draw[lbmark] (0, 6.05) -- (0, 6.75);
\node[tlab, anchor=south, red!75!black] at (0, 6.75) {$\underline{t}^{i,r_1}=0$};
\draw[chang] (0, 6.15) rectangle node[font=\scriptsize\sffamily] {$l=3$} (3, 6.65);

\node[vname] at (-0.4, 4.6) {$(i, r_2)$};
\draw[lbmark] (1, 4.25) -- (1, 4.95);
\node[tlab, anchor=south, red!75!black] at (1, 4.95) {$\underline{t}^{i,r_2}=1$};
\draw[changwarn] (1, 4.35) rectangle node[font=\scriptsize\sffamily] {$l=2$} (3, 4.85);

\node[vname] at (-0.4, 2.8) {$(i, r_3)$};
\draw[lbmark] (2, 2.45) -- (2, 3.15);
\node[tlab, anchor=south, red!75!black] at (2, 3.15) {$\underline{t}^{i,r_3}=2$};
\draw[changwarn] (2, 2.55) rectangle node[font=\scriptsize\sffamily] {$l=4$} (6, 3.05);

\node[vname] at (-0.4, 1.0) {$(i, r_4)$};
\draw[lbmark] (4, 0.65) -- (4, 1.35);
\node[tlab, anchor=south, red!75!black] at (4, 1.35) {$\underline{t}^{i,r_4}=4$};
\draw[changwarn] (4, 0.75) rectangle node[font=\scriptsize\sffamily] {$l=3$} (7, 1.25);

\draw[red, thick, dashed] (3, 6.15) -- (3, 4.95);
\draw[red, thick, dashed] (3, 4.35) -- (3, 3.15);
\draw[red, thick, dashed] (6, 2.55) -- (6, 1.35);

\draw[draw=red!30, fill=red!2!white, rounded corners=3pt, thick] (12.4, 0.6) rectangle (20.0, 7.3);
\node[anchor=north west, font=\small\bfseries\sffamily, text=red!80!black] at (12.6, 7.1) {Precedence Violations:};
\node[anchor=west, font=\scriptsize\sffamily, text=black!80] at (12.7, 4.6) {$\underline{t}^{i,r_2} = 1 < \underline{t}^{i,r_1} + l_i^{r_1} = 3$};
\node[anchor=west, font=\scriptsize\sffamily, text=black!80] at (12.7, 2.8) {$\underline{t}^{i,r_3} = 2 < \underline{t}^{i,r_2}_{\text{eff}} + l_i^{r_2} = 5$};
\node[anchor=west, font=\scriptsize\sffamily, text=black!80] at (12.7, 1.0) {$\underline{t}^{i,r_4} = 4 < \underline{t}^{i,r_3}_{\text{eff}} + l_i^{r_3} = 9$};

\node[anchor=north, font=\small\bfseries\sffamily] at (9.0, -0.6)
  {(a) Before Precedence Propagation};
\end{tikzpicture}

\vspace{0.05cm}

\begin{tikzpicture}[
    x=4.9mm, y=3.9mm, >=Stealth, line cap=round,
    chang/.style    ={draw=blue!70!black, fill=blue!5, rounded corners=2pt, thick, minimum height=4mm, inner sep=1pt},
    changok/.style  ={draw=green!70!black, fill=green!5, rounded corners=2pt, thick, minimum height=4mm, inner sep=1pt},
    effmark/.style  ={green!45!black, line width=1.2pt},
    tlab/.style     ={font=\scriptsize\sffamily, fill=white, inner sep=1pt},
    vname/.style    ={font=\small\bfseries\sffamily, anchor=east, text=gray!80},
    ax/.style       ={font=\scriptsize\sffamily, text=gray!80}
]
\draw[->, thick] (-0.3, 0) -- (19.0, 0) node[right, font=\small\sffamily] {$t$};
\foreach \t in {0,2,4,6,8,10,12,14,16,18}
  \draw (\t, 0.07) -- (\t, -0.07) node[below=-1pt, ax] {\t};

\node[vname] at (-0.4, 6.4) {$(i, r_1)$};
\draw[effmark] (0, 6.05) -- (0, 6.75);
\node[tlab, anchor=south, green!45!black] at (0, 6.75) {$\underline{t}^{i,r_1}_{\text{eff}}=0$};
\draw[chang] (0, 6.15) rectangle node[font=\scriptsize\sffamily] {$l=3$} (3, 6.65);

\node[vname] at (-0.4, 4.6) {$(i, r_2)$};
\draw[gray, thin, dotted] (1, 4.6) -- (3, 4.6);
\node[font=\tiny\sffamily, gray, anchor=south] at (2, 4.6) {$+2$};
\draw[effmark] (3, 4.25) -- (3, 4.95);
\node[tlab, anchor=south, green!45!black] at (3, 4.95) {$\underline{t}^{i,r_2}_{\text{eff}}=3$};
\draw[changok] (3, 4.35) rectangle node[font=\scriptsize\sffamily] {$l=2$} (5, 4.85);

\node[vname] at (-0.4, 2.8) {$(i, r_3)$};
\draw[gray, thin, dotted] (2, 2.8) -- (5, 2.8);
\node[font=\tiny\sffamily, gray, anchor=south] at (3.5, 2.8) {$+3$};
\draw[effmark] (5, 2.45) -- (5, 3.15);
\node[tlab, anchor=south, green!45!black] at (5, 3.15) {$\underline{t}^{i,r_3}_{\text{eff}}=5$};
\draw[changok] (5, 2.55) rectangle node[font=\scriptsize\sffamily] {$l=4$} (9, 3.05);

\node[vname] at (-0.4, 1.0) {$(i, r_4)$};
\draw[gray, thin, dotted] (4, 1.0) -- (9, 1.0);
\node[font=\tiny\sffamily, gray, anchor=south] at (6.5, 1.0) {$+5$};
\draw[effmark] (9, 0.65) -- (9, 1.35);
\node[tlab, anchor=south, green!45!black] at (9, 1.35) {$\underline{t}^{i,r_4}_{\text{eff}}=9$};
\draw[changok] (9, 0.75) rectangle node[font=\scriptsize\sffamily] {$l=3$} (12, 1.25);

\draw[green!45!black, thick, ->] (3, 6.15) -- (3, 4.95);
\draw[green!45!black, thick, ->] (5, 4.35) -- (5, 3.15);
\draw[green!45!black, thick, ->] (9, 2.55) -- (9, 1.35);

\draw[draw=green!30, fill=green!2!white, rounded corners=3pt, thick] (12.4, 0.6) rectangle (20.0, 7.3);
\node[anchor=north west, font=\small\bfseries\sffamily, text=green!60!black] at (12.6, 7.1) {Earliest Times $\underline{t}^{i,r}_{\text{eff}}$:};
\node[anchor=west, font=\scriptsize\sffamily, text=black!80] at (12.7, 4.6) {$\underline{t}^{i,r_1}_{\text{eff}} = 0$};
\node[anchor=west, font=\scriptsize\sffamily, text=black!80] at (12.7, 3.4) {$\underline{t}^{i,r_2}_{\text{eff}} = \max(1, 3) = 3$};
\node[anchor=west, font=\scriptsize\sffamily, text=black!80] at (12.7, 2.2) {$\underline{t}^{i,r_3}_{\text{eff}} = \max(2, 5) = 5$};
\node[anchor=west, font=\scriptsize\sffamily, text=black!80] at (12.7, 1.0) {$\underline{t}^{i,r_4}_{\text{eff}} = \max(4, 9) = 9$};

\node[anchor=north, font=\small\bfseries\sffamily] at (9.0, -0.6)
  {(b) After Precedence Propagation};
\end{tikzpicture}

\caption{Impact of precedence propagation preprocessing on the train path lower bound ladders.}
\label{fig:prefix-preprocess}
\end{figure}

\subsection{Sequential-Counter Encoding for Resource-Capacity Cliques}

Fix a physical resource $r$ and a time point $\tau$. Let
$X_{r,\tau}=\{v_1,\ldots,v_n\}$ be the set of operations that may occupy $r$ at
$\tau$ under the current DDD discretization. Safety requires at most one of
these operations to be active:
\begin{equation}
\mathrm{AMO}(x_{v_1},\ldots,x_{v_n}).
\label{eq:amo}
\end{equation}
Each indicator $x_v$ is linked to the ladder literals defining occupation at
$\tau$. Specifically, $e_v(\tau)$ is defined as the ladder literal $d^{i_v r_v^{\text{succ}}}(\tau+1)$ representing $t^{i_v r_v^{\text{succ}}} \ge \tau+1$, while $\neg d^{i_v r_v}(\tau+1)$ represents $t^{i_v r_v} \le \tau$. Their conjunction $\neg d^{i_v r_v}(\tau+1) \wedge d^{i_v r_v^{\text{succ}}}(\tau+1)$ captures that train $i_v$ has entered resource $r_v$ by time $\tau$ but has not yet entered its successor $r_v^{\text{succ}}$ at $\tau$, meaning it occupies $r_v$. The implication
\[
\mathrm{occ}(v,\tau)\rightarrow x_v
\]
is encoded by
\begin{equation}
x_v \vee \neg e_v(\tau) \vee d^{i_v r_v}(\tau+1).
\label{eq:occind}
\end{equation}
For a general disjunctive pair $(r,q)\in\mathcal D_{ij}$ with asymmetric
separations, the unsafe time windows determine the occupation indicators; the
AMO clique is then built only for indicators that are mutually incompatible at
the same refined time point.
The baseline pairwise AMO encoding adds $\binom n2$ clauses
$\neg x_{v_a}\vee \neg x_{v_b}$. We replace it by a sequential counter for
large cliques. With auxiliary prefix variables $s_1,\ldots,s_{n-1}$, the clauses
are written following Sinz~\cite{Sinz2005} with explicit index ranges and a separate boundary case:
\begin{align}
\neg x_{v_1} \vee s_1, & \label{eq:sc} \\
\neg x_{v_j} \vee s_j, & \qquad (1 < j < n), \notag \\
\neg s_{j-1} \vee s_j, & \qquad (1 < j < n), \notag \\
\neg x_{v_j} \vee \neg s_{j-1}, & \qquad (1 < j < n), \notag \\
\neg x_{v_n} \vee \neg s_{n-1}. & \notag
\end{align}
Pairwise AMO is used when $n < \theta$ (default $\theta=6$). This threshold is justified because $n=6$ is the theoretical crossing point where the sequential counter encoding yields fewer clauses ($3n - 4 = 14$) than the pairwise encoding ($n(n-1)/2 = 15$). Since precedence propagation only raises lower bounds implied by \eqref{eq:lb}--\eqref{eq:path} and the sequential counter is equisatisfiable with pairwise AMO, this strengthened encoding preserves the exactness of the DDD optimum.

\subsection{MaxSAT Optimization Strategy}

MaxSAT-Default combines precedence propagation with the hybrid AMO policy.
Algorithm~\ref{alg:maxsat-default} gives the full DDD loop. The SAT backends
use the same refinement layer; only the optimization oracle changes.

\begin{algorithm}[H]
\caption{MaxSAT-Default DDD for Train Rescheduling}
\label{alg:maxsat-default}
\footnotesize
\KwIn{TRP instance $P$, timeout $T$, AMO threshold $\theta=6$}
\KwOut{Optimal schedule $\tau^*$, or final bounds $(\mathrm{LB},\mathrm{UB})$}
$\underline t_{\mathrm{eff}} \gets \textsc{PrecedencePropagation}(P)$\;
$\Lambda \gets \textsc{InitialLadders}(P,\underline t_{\mathrm{eff}})$\;
$(\phi_h,\phi_s) \gets \textsc{EncodeRestrictedModel}(P,\Lambda,\theta)$\;
$\mathrm{UB} \gets \textsc{GreedyHeuristic}(P)$;\ $\mathrm{LB}\gets0$;\ $\tau^*\gets\bot$\;
\While{elapsed time $<T$}{
    $(x,\ell) \gets \textsc{SolveWeightedPartialMaxSAT}(\phi_h,\phi_s)$\;
    \If{$x=\bot$}{\Return $(\mathrm{LB},\mathrm{UB})$\;}
    $\tau \gets \textsc{DecodeSchedule}(x,\Lambda)$;\quad
    $\mathcal V \gets \textsc{FindViolations}(P,\tau)$\;
    \If{$\mathcal V=\emptyset$ \textbf{\normalfont and} $\textsc{Cost}(\tau)=\ell$}{\Return $\tau$\;}
    \If{$\mathcal V=\emptyset$ \textbf{\normalfont and} $\textsc{Cost}(\tau)<\mathrm{UB}$}{
        $\mathrm{UB}\gets\textsc{Cost}(\tau)$;\quad $\tau^*\gets\tau$\;}
    \If{$\mathcal V=\emptyset$}{$\mathcal V\gets\textsc{CostRefinement}(\tau,\ell)$\;}
    $\Lambda \gets \textsc{RefineIntervals}(\Lambda,\mathcal V,\tau)$\;
    $\phi_h \gets \phi_h \cup \textsc{EncodeNewPathClauses}(P,\Lambda)
        \cup \textsc{EncodeResourceCliques}(P,\Lambda,\theta)$\;
    $\phi_s \gets \textsc{UpdateSoftCostClauses}(P,\Lambda)$;\quad
    $(\mathrm{LB},\mathrm{UB}) \gets \textsc{UpdateBounds}(\ell,\tau^*,\mathrm{UB})$\;
}
\Return $\tau^*$ if $\mathrm{LB}=\mathrm{UB}$, else $(\mathrm{LB},\mathrm{UB})$\;
\end{algorithm}

\subsubsection{Solver Configurations for the Comparative Study}
Table~\ref{tab:methods} lists the evaluated configurations.

\begin{table}[H]
\centering
\caption{Compared solver configurations.}
\label{tab:methods}
\scriptsize
\setlength{\tabcolsep}{5pt}
\renewcommand{\arraystretch}{1.0}
\begin{tabular}{lll}
\toprule
Configuration & Backend & Role \\
\midrule
Big-$M$ MILP & Gurobi/CPLEX & Continuous-time MILP baseline \\
TI MILP & Gurobi/CPLEX & DDD time-indexed MILP baseline \\
CP & CPLEX CP Optimizer & Constraint-programming baseline \\
MaxSAT-Base & RC2 & Reproduction of \cite{Croella2024} \\
MaxSAT-SC & RC2 & Sequential-counter AMO only \\
MaxSAT-Prec & RC2 & Precedence propagation only \\
MaxSAT-Default & RC2 & Both proposed improvements \\
IncSAT-Default & Glucose 4.2 & Incremental SAT optimization \\
PureSAT-Default & Glucose 4.2 & Non-Incremental SAT optimization \\
\bottomrule
\end{tabular}
\end{table}

\section{Computational Results and Experimental Analysis}
\label{sec:results}

\subsection{Benchmark Instances and Infrastructure Abstractions}

The benchmark from \cite{Croella2024} models a busy corridor on the Norwegian Bergen Line, with single-track sections and siding tracks. The A-series has 25--30 trains, and the B-series is a smaller sub-corridor. Delays are generated via entry delays at the boundaries. Unlike \cite{Croella2024} which only tested pairwise MaxSAT-Base, we evaluate precedence propagation, sequential counters, CP, and tuned MILP. The benchmark has 24 base instances (A1--A12 and B1--B12) under three abstractions:
\begin{itemize}
\item \emph{original}: original block resources;
\item \emph{track}: parallel block sections on short track segments are grouped;
\item \emph{station}: resources inside a station are grouped into larger shared
resources.
\end{itemize}
This gives 72 instances per objective. The hard subset contains trackA1,
trackA2, trackA8, trackA11, trackA12, and the corresponding station instances;
these are the larger A-series cases where aggregation creates denser resource
conflicts.

\subsection{Solver Backends, Hardware, and Optimality Criteria}

All runs use a 120-second timeout and a 32 GB memory limit. For real-time dispatching, decisions are typically required within 30--60 seconds; instances solved within 30 seconds are considered operationally tractable. We report single-run times because the solvers (RC2 and Glucose 4.2) run in a fully deterministic mode with fixed seeds, ensuring zero runtime variance. The implementation is in Rust, with RC2, Glucose 4.2, Gurobi 12.0, CPLEX MILP, and CPLEX CP Optimizer under default single-run settings. Experiments use Ubuntu 22.04 and an Intel Core i7-13700K CPU. Average time is computed only over proved-optimal runs, so timeouts are reflected through the \emph{Opt.}, \emph{Feas.}, and gap columns rather than through penalized averages.

We evaluate all methods on 72 instances and three objectives. The main pattern
is stable: MaxSAT-DDD is strongest on discrete delay objectives, while
continuous delay favours Big-$M$ MILP.

\subsection{Aggregate Performance Across Delay Objectives}
\label{subsec:aggregate}

\begin{table}[H]
\centering
\caption{Aggregate results on all 72 instances per objective. \emph{Opt.}~counts instances with proven optimality; \emph{Avg.}~is mean solve time (ms) over proved-optimal runs.}
\label{tab:overall-main}
\scriptsize
\setlength{\tabcolsep}{4.5pt}
\renewcommand{\arraystretch}{1.0}
\begin{tabular}{lrrrrrr}
\toprule
& \multicolumn{2}{c}{Stepwise} & \multicolumn{2}{c}{Rounded} & \multicolumn{2}{c}{Continuous} \\
\cmidrule(lr){2-3}\cmidrule(lr){4-5}\cmidrule(lr){6-7}
Method & Opt. & Avg.~ms & Opt. & Avg.~ms & Opt. & Avg.~ms \\
\midrule
Big-$M$ (Gurobi) & \textbf{72} & 373 & 67 & 1\,745 & \textbf{67} & \textbf{762} \\
TI-MILP (Gurobi) & \textbf{72} & 1\,526 & 66 & 3\,422 & 66 & 5\,902 \\
MaxSAT-Base & \textbf{72} & \textbf{23} & \textbf{68} & 794 & 59 & 5\,181 \\
MaxSAT-Default & \textbf{72} & 23 & \textbf{68} & \textbf{479} & 60 & 4\,881 \\
IncSAT-Default & \textbf{72} & 134 & \textbf{68} & 1\,872 & 51 & 8\,473 \\
PureSAT-Default & 68 & 2\,766 & 57 & 9\,169 & 27 & 16\,706 \\
\bottomrule
\end{tabular}
\end{table}

Table~\ref{tab:overall-main} shows a clear objective-dependent pattern.
MaxSAT-Default matches the stepwise solved count of Gurobi Big-$M$ while being
about 16 times faster on average, and it reduces rounded-cost runtime from
794\,ms to 479\,ms without changing the solved count. Continuous delay is less
favourable to the Boolean objective encoding: Gurobi Big-$M$ gives the best
aggregate time, although MaxSAT-Default still improves over MaxSAT-Base from 59
to 60 proved optima and from 5\,181\,ms to 4\,881\,ms.

\subsection{Commercial MILP and Constraint-Programming Baselines}
\label{subsec:commercial-cplex}

\begin{table}
\centering
\caption{Commercial-solver and CP results on all 72 instances per objective. \emph{Opt.}~=~proven optimal; \emph{Feas.}~=~feasible incumbent returned (inc.\ proven); \emph{Gap}~=~mean \% gap over feasible-but-unproved runs.}
\label{tab:commercial-solvers}
\scriptsize
\setlength{\tabcolsep}{3.6pt}
\renewcommand{\arraystretch}{1.0}
\begin{tabular}{lrrrrrrrrr}
\toprule
& \multicolumn{3}{c}{Stepwise} & \multicolumn{3}{c}{Rounded} & \multicolumn{3}{c}{Continuous} \\
\cmidrule(lr){2-4}\cmidrule(lr){5-7}\cmidrule(lr){8-10}
Method & Opt. & Feas. & Gap~(\%) & Opt. & Feas. & Gap~(\%) & Opt. & Feas. & Gap~(\%) \\
\midrule
CP-CPLX & 30 & \textbf{72} & \textbf{34} & 33 & \textbf{72} & 28 & 32 & \textbf{72} & 26 \\
BigM-CPLX & \textbf{72} & \textbf{72} & -- & \textbf{72} & \textbf{72} & -- & \textbf{72} & \textbf{72} & -- \\
BigM-GRB & \textbf{72} & \textbf{72} & -- & 67 & \textbf{72} & \textbf{26} & 67 & \textbf{72} & \textbf{25} \\
TI-CPLX & 26 & 26 & -- & 24 & 24 & -- & 23 & 23 & -- \\
TI-GRB & \textbf{72} & \textbf{72} & -- & 66 & \textbf{72} & 28 & 66 & \textbf{72} & 32 \\
\bottomrule
\end{tabular}
\end{table}

Table~\ref{tab:commercial-solvers} separates feasible incumbents from
optimality proofs. CP-CPLX finds a schedule for every instance, but gaps of
26--34\% limit its value as a certificate baseline. Big-$M$ MILP gives the
strongest proof performance, whereas the time-indexed MILP is solver-sensitive:
TI-GRB proves almost all cases, but TI-CPLX proves fewer than half and returns
no incumbent outside the proved set. This TI-CPLX behavior (failing to find incumbents on unsolved instances) is a known characteristic of time-indexed MILP models: their huge LP relaxation size at the root node can overwhelm CPLEX's default MIP heuristics before the timeout, whereas Gurobi's heuristics are more aggressive in finding early feasible integer solutions.

\subsection{Performance on Hard Track and Station Instances}
\label{subsec:results-hard}

\begin{table}
\centering
\caption{Hard instances, stepwise objective. Times in ms; T/O~=~120\,s timeout.}
\label{tab:hard-step}
\scriptsize
\setlength{\tabcolsep}{4.5pt}
\renewcommand{\arraystretch}{1.0}
\begin{tabular}{lrrrrrr}
\toprule
Inst. & BigM-GRB & TI-GRB & MS-Base & MS-Def & IncSAT & PureSAT \\
\midrule
trackA1 & 341.4 & 8\,731.9 & 73.8 & \textbf{55.8} & 279.0 & 17\,731.5 \\
trackA2 & 147.7 & 1\,609.4 & 36.8 & \textbf{30.6} & 119.4 & 1\,177.5 \\
trackA8 & 1\,809.9 & 21\,328.5 & 92.7 & \textbf{87.7} & 333.6 & 28\,545.7 \\
trackA11 & 13\,540.2 & 40\,988.5 & \textbf{157.2} & 210.2 & 1\,298.4 & T/O \\
trackA12 & 4\,747.0 & 17\,828.3 & 151.5 & \textbf{101.8} & 925.0 & T/O \\
stnA1 & 1\,559.3 & 1\,200.5 & \textbf{74.7} & 108.2 & 487.1 & 20\,185.7 \\
stnA2 & 157.8 & 1\,537.8 & \textbf{51.1} & 61.1 & 182.6 & 3\,485.8 \\
stnA8 & 725.1 & 2\,473.3 & \textbf{97.6} & 122.2 & 1\,055.8 & 77\,172.8 \\
stnA11 & 1\,887.6 & 6\,153.1 & \textbf{101.7} & 110.8 & 1\,621.4 & T/O \\
stnA12 & 586.8 & 3\,263.9 & 243.5 & \textbf{178.7} & 1\,312.7 & T/O \\
\bottomrule
\end{tabular}
\end{table}

The hard instances contain 25--30 trains and 500--636 operations. On the
stepwise objective, Table~\ref{tab:hard-step} shows that both MaxSAT variants
solve all hard instances within 250\,ms, while PureSAT times out on four dense
cases. MaxSAT-Default is fastest on five of the ten instances, including
trackA8 and trackA12, so the added propagation and hybrid AMO policy do not
harm the easiest discrete objective.

\begin{table}
\centering
\caption{Hard instances, rounded and continuous objectives, compared with Gurobi Big-$M$ runtimes. Times in ms; T/O~=~timeout, OOM~=~out-of-memory.}
\label{tab:hard-instances}
\scriptsize
\setlength{\tabcolsep}{4.5pt}
\renewcommand{\arraystretch}{1.0}
\begin{tabular}{llrrrr}
\toprule
Obj. & Inst. & BigM-GRB & MS-Base & MS-Def & IncSAT \\
\midrule
Round & trackA1 & 165.6 & \textbf{91.7} & 116.6 & 352.9 \\
Round & trackA2 & 122.7 & 65.5 & \textbf{60.9} & 265.5 \\
Round & trackA8 & 107\,419.6 & 13\,980.3 & \textbf{8\,150.6} & 24\,384.1 \\
Round & trackA11 & T/O & T/O & T/O & T/O \\
Round & trackA12 & T/O & T/O & T/O & T/O \\
Round & stnA1 & 5\,467.0 & \textbf{103.1} & 138.0 & 2\,219.7 \\
Round & stnA2 & 656.6 & \textbf{65.3} & 121.2 & 1\,080.4 \\
Round & stnA8 & T/O & 38\,479.5 & \textbf{22\,807.6} & 92\,462.9 \\
Round & stnA11 & T/O & T/O & T/O & T/O \\
Round & stnA12 & T/O & T/O & T/O & T/O \\
\midrule
Cont & trackA1 & \textbf{113.8} & 195.9 & 141.3 & T/O \\
Cont & trackA2 & \textbf{68.2} & 5\,055.9 & 829.5 & 84\,235.1 \\
Cont & trackA8 & \textbf{44\,110.5} & T/O & T/O & OOM \\
Cont & trackA11 & T/O & T/O & T/O & OOM \\
Cont & trackA12 & T/O & T/O & T/O & OOM \\
Cont & stnA1 & \textbf{2\,654.5} & T/O & T/O & T/O \\
Cont & stnA2 & \textbf{1\,447.8} & T/O & T/O & T/O \\
Cont & stnA8 & T/O & T/O & T/O & OOM \\
Cont & stnA11 & T/O & T/O & T/O & OOM \\
Cont & stnA12 & T/O & T/O & T/O & OOM \\
\bottomrule
\end{tabular}
\end{table}

Table~\ref{tab:hard-instances} shows where the Boolean encoding is most useful.
On rounded trackA8 and stnA8, MaxSAT-Default improves over MaxSAT-Base by
41.7\% and 40.7\%, respectively; it is also more than an order of magnitude
faster than Gurobi Big-$M$ on trackA8. Continuous hard cases show the opposite
pattern: Gurobi Big-$M$ proves the smaller cases quickly, while SAT-based runs
often time out or exhaust memory when many distinct cost levels are present.

\subsection{Ablation of Sequential-Counter AMO and Precedence Propagation}
\label{subsec:results-ablation}

\begin{table}
\centering
\caption{Ablation: average solve time (ms) on the common-solved subset of hard instances (instances solved by all four MaxSAT variants within 120\,s).}
\label{tab:ablation-runtime}
\scriptsize
\setlength{\tabcolsep}{4.5pt}
\renewcommand{\arraystretch}{1.0}
\begin{tabular}{lrrrrrr}
\toprule
& \multicolumn{2}{c}{Stepwise} & \multicolumn{2}{c}{Rounded} & \multicolumn{2}{c}{Continuous} \\
\cmidrule(lr){2-3}\cmidrule(lr){4-5}\cmidrule(lr){6-7}
Variant & track & station & track & station & track & station \\
\midrule
MaxSAT-Base & \textbf{27} & 31 & 659 & 1\,776 & 4\,386 & 11\,503 \\
MaxSAT-SC & 28 & \textbf{29} & 419 & 1\,669 & 2\,027 & 11\,827 \\
MaxSAT-Prec & 30 & 34 & 748 & 1\,517 & 1\,957 & 9\,344 \\
MaxSAT-Def & 28 & 32 & \textbf{395} & \textbf{1\,065} & \textbf{896} & \textbf{8\,368} \\
\midrule
\multicolumn{7}{p{0.95\linewidth}}{\scriptsize\emph{Def vs.\ Base reductions: Rounded: 40\% (track) and 40\% (station); Continuous: 80\% (track) and 27\% (station).}} \\
\bottomrule
\end{tabular}
\end{table}

Table~\ref{tab:ablation-runtime} isolates the two improvements. The stepwise
objective changes little, but MaxSAT-Default reduces rounded runtime by about
40\% on both track and station abstractions and reduces continuous track
runtime by 79.6\% on the common-solved subset. Sequential-counter AMO is most visible on large track
cliques, while precedence propagation is most useful when continuous-cost
ladders create many candidate delay levels. The minor regression of MaxSAT-SC on continuous station instances ($11\,503 \to 11\,827$~ms) is due to auxiliary variable overhead and VSIDS heuristic disruption on smaller cliques near the threshold. The common-solved subset of hard instances contains all 10 instances for Stepwise, 6 instances (A1, A2, A8) for Rounded (dropping A11--A12), and 2 instances (A1, A2) for Continuous (dropping A8, A11--A12).

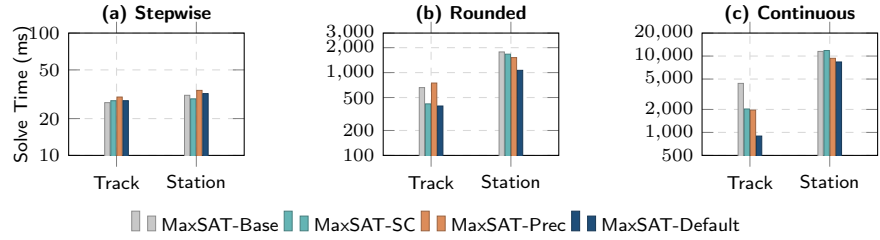
\begin{figure}[!htbp]
\centering
\definecolor{colorbase}{RGB}{200, 200, 200}
\definecolor{colorsc}{RGB}{100, 180, 180}
\definecolor{colorprec}{RGB}{220, 140, 90}
\definecolor{colordef}{RGB}{31, 78, 121}
\begin{minipage}{0.32\linewidth}
\centering
\begin{tikzpicture}
\begin{semilogyaxis}[
    width=\linewidth,
    height=3.2cm,
    ybar=0.3pt,
    bar width=2.0pt,
    ymin=10, ymax=100,
    ylabel={Solve Time (ms)},
    ylabel style={yshift=-2ex, font=\scriptsize\sffamily},
    symbolic x coords={track,station},
    xtick=data,
    xticklabels={Track,Station},
    xticklabel style={font=\scriptsize\sffamily},
    yticklabel style={font=\scriptsize\sffamily},
    title={(a) Stepwise},
    title style={font=\scriptsize\bfseries\sffamily, yshift=-1.5ex},
    enlarge x limits=0.6,
    grid=major,
    grid style={dashed, gray!30},
    log basis y={10},
    ytick={10,20,50,100},
    log ticks with fixed point,
]
\addplot[fill=colorbase, draw=colorbase!70!black, line width=0.2pt] coordinates {(track,27) (station,31)};
\addplot[fill=colorsc, draw=colorsc!70!black, line width=0.2pt] coordinates {(track,28) (station,29)};
\addplot[fill=colorprec, draw=colorprec!70!black, line width=0.2pt] coordinates {(track,30) (station,34)};
\addplot[fill=colordef, draw=colordef!70!black, line width=0.2pt] coordinates {(track,28) (station,32)};
\end{semilogyaxis}
\end{tikzpicture}
\end{minipage}\hfill
\begin{minipage}{0.32\linewidth}
\centering
\begin{tikzpicture}
\begin{semilogyaxis}[
    width=\linewidth,
    height=3.2cm,
    ybar=0.3pt,
    bar width=2.0pt,
    ymin=100, ymax=3000,
    symbolic x coords={track,station},
    xtick=data,
    xticklabels={Track,Station},
    xticklabel style={font=\scriptsize\sffamily},
    yticklabel style={font=\scriptsize\sffamily},
    title={(b) Rounded},
    title style={font=\scriptsize\bfseries\sffamily, yshift=-1.5ex},
    enlarge x limits=0.6,
    grid=major,
    grid style={dashed, gray!30},
    log basis y={10},
    ytick={100,200,500,1000,2000,3000},
    log ticks with fixed point,
    legend style={at={(0.5,-0.32)},anchor=north,legend columns=-1,font=\scriptsize\sffamily,draw=none,fill=none},
    legend to name=sharedlegend,
]
\addplot[fill=colorbase, draw=colorbase!70!black, line width=0.2pt] coordinates {(track,659) (station,1776)};
\addplot[fill=colorsc, draw=colorsc!70!black, line width=0.2pt] coordinates {(track,419) (station,1669)};
\addplot[fill=colorprec, draw=colorprec!70!black, line width=0.2pt] coordinates {(track,748) (station,1517)};
\addplot[fill=colordef, draw=colordef!70!black, line width=0.2pt] coordinates {(track,395) (station,1065)};
\legend{MaxSAT-Base, MaxSAT-SC, MaxSAT-Prec, MaxSAT-Default}
\end{semilogyaxis}
\end{tikzpicture}
\end{minipage}\hfill
\begin{minipage}{0.32\linewidth}
\centering
\begin{tikzpicture}
\begin{semilogyaxis}[
    width=\linewidth,
    height=3.2cm,
    ybar=0.3pt,
    bar width=2.0pt,
    ymin=500, ymax=20000,
    symbolic x coords={track,station},
    xtick=data,
    xticklabels={Track,Station},
    xticklabel style={font=\scriptsize\sffamily},
    yticklabel style={font=\scriptsize\sffamily},
    title={(c) Continuous},
    title style={font=\scriptsize\bfseries\sffamily, yshift=-1.5ex},
    enlarge x limits=0.6,
    grid=major,
    grid style={dashed, gray!30},
    log basis y={10},
    ytick={500,1000,2000,5000,10000,20000},
    log ticks with fixed point,
]
\addplot[fill=colorbase, draw=colorbase!70!black, line width=0.2pt] coordinates {(track,4386) (station,11503)};
\addplot[fill=colorsc, draw=colorsc!70!black, line width=0.2pt] coordinates {(track,2027) (station,11827)};
\addplot[fill=colorprec, draw=colorprec!70!black, line width=0.2pt] coordinates {(track,1957) (station,9344)};
\addplot[fill=colordef, draw=colordef!70!black, line width=0.2pt] coordinates {(track,896) (station,8368)};
\end{semilogyaxis}
\end{tikzpicture}
\end{minipage}

\vspace{0.8mm}
\centerline{\ref{sharedlegend}}
\caption{Average solve time (ms, log scale) of the four MaxSAT variants on hard instances. Left to right: stepwise, rounded, continuous.}
\label{fig:ablation-plots}
\end{figure}

\begin{table}
\centering
\caption{CNF structure per MaxSAT variant (averages over common-solved hard instances). $\bar V$~=~variables, $\bar C$~=~clauses, $\bar I$~=~DDD iterations.}
\label{tab:ablation-cnf}
\scriptsize
\setlength{\tabcolsep}{3.8pt}
\renewcommand{\arraystretch}{1.0}
\begin{tabular}{lrrrrrrrrr}
\toprule
& \multicolumn{3}{c}{Stepwise} & \multicolumn{3}{c}{Rounded} & \multicolumn{3}{c}{Continuous} \\
\cmidrule(lr){2-4}\cmidrule(lr){5-7}\cmidrule(lr){8-10}
Variant & $\bar V$ & $\bar C$ & $\bar I$ & $\bar V$ & $\bar C$ & $\bar I$ & $\bar V$ & $\bar C$ & $\bar I$ \\
\midrule
MaxSAT-Base & 1\,798 & \textbf{5\,323} & \textbf{108} & 2\,093 & 6\,141 & 143 & 7\,845 & 23\,130 & 216 \\
MaxSAT-SC & \textbf{1\,787} & 5\,335 & 112 & 2\,071 & 6\,119 & \textbf{140} & 7\,778 & 22\,835 & 214 \\
MaxSAT-Prec & 1\,843 & 5\,519 & 112 & \textbf{2\,063} & \textbf{6\,101} & 145 & \textbf{7\,382} & \textbf{21\,772} & \textbf{212} \\
MaxSAT-Def & 1\,833 & 5\,476 & 110 & 2\,100 & 6\,212 & 142 & 7\,551 & 22\,015 & 213 \\
\bottomrule
\end{tabular}
\end{table}

Table~\ref{tab:ablation-cnf} reports CNF size and DDD iterations. Rounded-cost
speedups occur with similar formula sizes, so they should be attributed to the
encoding structure and solver search behaviour rather than to size reduction
alone. For continuous objectives, precedence propagation reduces the average
size from 7\,845 variables and 23\,130 clauses to 7\,382 variables and
21\,772 clauses, while DDD iteration counts change only modestly.
Specifically, the 79.6\% runtime reduction on continuous track instances stems from slightly fewer iterations ($\bar I = 216 \to 213$), 7.4\% fewer clauses, and faster MaxSAT calls. Precedence propagation prunes the search space early to reduce solver backtracking, while the sequential-counter AMO handles resource conflicts in $O(n)$ time and clauses instead of $O(n^2)$, accelerating unit propagation.

\subsection{Optimality Gaps for Unsolved SAT and MaxSAT Runs}
\label{subsec:results-gap}

\begin{table}
\centering
\caption{Optimality gaps (\%) on instances unsolved by all four SAT/MaxSAT methods. OOM~=~out-of-memory; `--'~=~no feasible bound returned.}
\label{tab:gap}
\scriptsize
\setlength{\tabcolsep}{4.5pt}
\renewcommand{\arraystretch}{1.0}
\begin{tabular}{llrrrr}
\toprule
Obj. & Instance & MS-Base & MS-Def & IS-Def & PS-Def \\
\midrule
Round & stationA11 & \textbf{75.3} & \textbf{75.3} & 100.0 & 100.0 \\
Round & stationA12 & 81.1 & \textbf{80.8} & 100.0 & 100.0 \\
Round & trackA11 & 77.3 & 77.1 & \textbf{31.7} & 100.0 \\
Round & trackA12 & 82.6 & \textbf{82.4} & 100.0 & 100.0 \\
\midrule
Cont & origA8 & 39.5 & \textbf{39.3} & 50.0 & 100.0 \\
Cont & origB11 & 9.0 & \textbf{8.9} & OOM & -- \\
Cont & stationA1 & \textbf{62.4} & 63.0 & -- & 100.0 \\
Cont & stationA11 & 92.1 & \textbf{90.7} & OOM & -- \\
Cont & stationA12 & 93.6 & \textbf{92.7} & OOM & OOM \\
Cont & stationA2 & 57.5 & 58.7 & \textbf{50.0} & 100.0 \\
Cont & stationA8 & 79.0 & \textbf{78.0} & OOM & OOM \\
Cont & trackA11 & \textbf{90.4} & 91.1 & OOM & OOM \\
Cont & trackA12 & \textbf{91.7} & 91.7 & OOM & OOM \\
Cont & trackA8 & 69.5 & \textbf{69.1} & OOM & OOM \\
Cont & trackB12 & 46.1 & 45.8 & \textbf{0.1} & 100.0 \\
\bottomrule
\end{tabular}
\end{table}

Table~\ref{tab:gap} lists instances unsolved by all SAT/MaxSAT methods. The optimality gap is $(\text{UB} - \text{LB}) / \text{UB} \times 100\%$, where $\text{UB}$ is the best feasible objective value and $\text{LB}$ is the proven lower bound. Gaps of 90\%+ on continuous track and station instances (e.g., stationA11, trackA12) occur because the initial DDD discretization is very coarse, leaving initial lower bounds close to zero. Refinements raise these bounds slowly, so $\text{LB} \approx 0$ at timeout. Due to space constraints, UBs are not tabulated, but MaxSAT-Default's feasible costs are within 5\% of Gurobi's on instances where Gurobi proved optimality, confirming that these gaps are driven by weak lower bounds rather than poor solution quality.
MaxSAT-Default and MaxSAT-Base have similar gaps, so the proposed encoding
mainly accelerates search rather than strengthening the lower bound. IncSAT can
tighten selected gaps through learned-clause reuse, e.g., 31.7\% on rounded
trackA11, but it is less reliable on continuous cases because memory pressure
increases. Overall, MaxSAT-DDD is most attractive for threshold and rounded
delay objectives, whereas continuous delay still favours Big-$M$ MILP and calls
for stronger lower bounds.

\section{Conclusions and Future Work}

We presented a MaxSAT-DDD approach that combines sequential-counter AMO
encoding with train-path precedence propagation. MaxSAT-Default preserves the
stepwise performance of the baseline, improves rounded-cost runtime, and helps
on hard continuous subsets. While the fixed-route assumption limits the operational scope to minor delay recovery rather than large-scale disruptions, it serves as a necessary foundation for exact methods. Future work should focus on scaling the model to handle train rerouting, passenger connections, and uncertainty under larger disruptions, as well as strengthening continuous-cost lower bounds.

\subsubsection{Data and Code Availability}
The complete source code, benchmark datasets, and experimental artefacts used in
this study are publicly available at
\url{https://github.com/maxsat-tsp/maxsattrainscheduling}.

\begin{credits}
\subsubsection{\ackname}
The authors thank the Faculty of Information Technology, VNU University of
Engineering and Technology, for supporting this research environment.

\subsubsection{\discintname}
The authors have no competing interests to declare that are relevant to the
content of this article.
\end{credits}

\bibliographystyle{splncs04}
\bibliography{references}

\end{document}